\documentclass[%
 reprint,
 amsmath,amssymb,
 aps,
]{revtex4-2}

\usepackage{graphicx}% Include figure files
\usepackage{dcolumn}% Align table columns on decimal point
\usepackage{bm}% bold math

\usepackage{type1cm}        % activate if the above 3 fonts are
\usepackage{makeidx}         % allows index generation
\usepackage[bottom]{footmisc}% places footnotes at page bottom

\usepackage{newtxtext}       % 
\usepackage{newtxmath}       % selects Times Roman as basic font

\usepackage{url}

\newcounter{inlineenum}
\renewcommand{\theinlineenum}{\roman{inlineenum}}
\newenvironment{inlineenum}
{\unskip\ignorespaces\setcounter{inlineenum}{0}%
	\renewcommand{\item}{\refstepcounter{inlineenum}{\textit{\theinlineenum})~}}}
{\ignorespacesafterend}

\begin{document}

\title{Collective traffic of agents that remember}% 
\thanks{Presented at the Traffic and Granular Flow 22 conference, held in New Delhi, India.}%

\author{Danny Raj M}
    \email{dannym@iisc.ac.in; dannyrajmasila@gmail.com}
\author{Arvind Nayak}%
\affiliation{%
Dept of Chemical Engineering, Indian Institute of Science Bangalore, Bengaluru 560012, Karnataka, India
}%

\date{\today}% It is always \today, today,

\begin{abstract}
Traffic and pedestrian systems consist of human collectives where agents are intelligent and capable of processing available information, to perform tactical manoeuvres that can potentially increase their movement efficiency.
In this study, we introduce a social force model for agents that possess \textit{memory}. Information of the agent's past affects the agent's instantaneous movement in order to swiftly take the agent towards its desired state.
We show how the presence of memory is akin to an agent performing a proportional--integral control to achieve its desired state.
% Depending on how long the agent remembers and the impact of memory on the motion of the agent, we find that a single isolated agent can show four qualitatively distinct dynamics.
The longer the agent remembers and the more impact the memory has on its motion, better is the movement of an isolated agent in terms of achieving its desired state.
However, when in a collective, the interactions between the agents lead to non-monotonic effect of memory on the traffic dynamics. A group of agents with memory exiting through a narrow door exhibit more clogging with memory than without it. We also show that a very large amount of memory results in variation in the memory force experienced by agents in the system at any time, which reduces the propensity to form clogs and leads to efficient movement.
\end{abstract}

\maketitle

\section{Introduction}
The ability to think is a key distinction between agents in human collectives, like vehicular and pedestrian traffic flows~\cite{Moussaid2009, Seitz2016, Rahmati2018, Nicolas2019}, and granular collectives~\cite{Candelier2010, Guo2015}, which are made up of grains that respond and move only to external forcing. Therefore, when one writes down a set of governing equations either in the form of rules of interaction or social forces, it is important to specify those that distinguish a living agent from one that is non-living. It would be an interesting problem to explore how this distinction manifests in the dynamics of the collective. 
Since a comprehensive model for human cognition (or thinking) is incredibly hard to formulate, we adopt a bottom-up approach where we investigate how a certain facet of intelligence affects the collective dynamics.

One of the key characteristics of a living agent is its ability to assimilate information, process it and alter the agent’s decisions based on it. A critical aspect in this process is the agent’s memory. An agent can remember its past actions and change its current actions accordingly to achieve better performance. Memory of an agent can affect the dynamics across several time scales: from learning route-specific rules over several days and months, to remembering its recent trajectory and reacting to immediate changes in the environment/jams. Here, we concentrate on the latter. We develop a social force model for the memory of an agent based on its past movement. Remembering how it moved in the recent past, the agent evaluates how well it has achieved its desired velocity and takes a tactical decision---to make up for the sub-optimal past movement. 

In this article, we introduce a social force model for agents that exhibit memory. Memory is characterised by two parameters:
\begin{inlineenum}
    \item how long the agent remembers and,
    \item impact of memory on the movement.
\end{inlineenum}
We first analyse the model equations to understand the impact of memory on the dynamics of an isolated agent. Then, we proceed to test the effect of memory on the collective. We simulate agents attempting to exit via narrow door where agents are known to exhibit clogging behaviour near the exits.
Our objective is to understand how memory aids in the collective motion: Does the presence of memory always guarantee efficient movement? How does it impact the propensity to form or displace agents in a clog?

\section{Model for traffic dynamics}
\subsection{Agents with memory}
The motion of agents is modelled using a social force model, similar to Helbing et al~\cite{Helbing1995, Helbing2000,Helbing2000a}. The velocity of an agent evolves in time based on Eq~\ref{eqn:socialforcemodel}, where: 
\begin{inlineenum}
    \item the first term in the RHS is the restitution force that restores the agent to its desired direction and speed $\mathbf{v}_0$,
    \item the second term is the force due to the effect of the memory of an agent and,
    \item the third term $\mathbf{I}_{j,b}^i$ is the net social interaction of the agent to avoid collision with other agents and the boundary.
\end{inlineenum}

\begin{equation}
    m\frac{d\mathbf{v}_i}{dt} =  \frac{m}{\tau} (\mathbf{v}_0 - \mathbf{v}_i) + \Tilde{\beta} \mathbf{M}_i + I_{j,b}^i
    \label{eqn:socialforcemodel}
\end{equation}

Here, the memory $\mathbf{M}_i(t)$ of an agent $i$ at time $t$, is defined as the total deviation of the velocity of the agent from its desired velocity $\mathbf{v}_0$ in the time window $[t-T, t]$ of the agent's recent past (see Eq~\ref{eqn:memoryterm}). 

\begin{equation}
    \mathbf{M}_i = \int_{t-T}^t \big(\mathbf{v}_0 - \mathbf{v}_i(t')\big) dt'
    \label{eqn:memoryterm}
\end{equation}

When the agent is unable to travel at its desired speed and direction, the memory term takes a positive value and offers an additional force to help restore the agent's motion sooner.

\begin{equation}
    I_{j,b}^i = \sum_{\forall j \in \mathcal{N}_i}\Big[\mathbf{F}_{i,j}^n + \mathbf{F}_{i,j}^t\Big] + \mathbf{F}_{b}^n + \mathbf{F}_{b}^t
    \label{eqn:agent2boundaryforces}
\end{equation}

The term $\mathbf{I}_{j,b}^i$ can take different forms depending on the context of the traffic problem. For pedestrian dynamics~\cite{Helbing2000}, Helbing and co-workers considered interaction forces similar to that used in granular flows: a sum of the total normal and tangential (frictional) forces due to contact. However, $\mathbf{I}_{j,b}^i$ can also include small-ranged forces that prevent collisions~\cite{Helbing2000a}, which are more suitable for traffic flow problems. The qualitative results shown in the paper do not depend on the exact choice of the models for interactions.

\subsection{Connections to control theory.}
If one were to imagine $\mathbf{v}_0$ as the set-point for a given agent, \textit{i.e.}, the desired \textit{state} that the agent wants to achieve, then the restitution force and the memory term in Eqs~\ref{eqn:socialforcemodel} and \ref{eqn:memoryterm}, exactly resemble a proportional and an integral parts of a controller (PI), respectively.
Presence of an integral component can result in overshoots and oscillations before the agent reaches its set-point. In the absence of memory and when there are continuous collisions with obstacles as agents move, we can expect the dynamics of the agent to show an offset (not reach $\|\mathbf{v}_0\|$)---since, it is well known that a proportional controller cannot take the system exactly to its set-point. Addition of memory could guarantee reaching the set-point, even in the presence of obstacles.

One could also conceive a dynamics resembling a PID controller. This modification does not qualitatively change the dynamics of the system. Since, adding a derivative term for the deviation $\mathbf{v}_0 - \mathbf{v}_i$, gets absorbed into the inertia (LHS term of Eq~\ref{eqn:socialforcemodel}). In other words, the derivative term acts like an effective \textit{mass} term that slows the response of the agent to any social force.

\subsection{Memory as a state of the agent.}
The memory term in Eq~\ref{eqn:memoryterm} is computed within a time window of $[t-T, t]$. Any information from outside the time window is not used and all the information within are equally weighted.
A simpler and more elegant formulation can be arrived at, if we weight the information with an exponential weighting as shown in Eq~\ref{eqn:memorywithExpweigthing}.
\begin{equation}
    \mathbf{M}_i = \int_{0}^t e^{\frac{t'-t}{\Tilde{\alpha}}} \big(\mathbf{v}_0 - \mathbf{v}_i(t')\big) dt'
    \label{eqn:memorywithExpweigthing}
\end{equation}
An exponential weighting prioritises information close to the current time instant than from a distant past.
This allows us to move away from a discontinuous time window set by $T$ and towards a time scale $\tilde{\alpha}$.
Then, we can differentiate Eq~\ref{eqn:memorywithExpweigthing} with time, using the Leibniz rule for differentiating under the integral sign to get Eq~\ref{eqn:MemoryExpModel}.
\begin{equation}
    \frac{d\mathbf{M}_i}{dt} = \mathbf{v}_0 - \mathbf{v}_i - \frac{\mathbf{M}_i}{\Tilde{\alpha}}
    \label{eqn:MemoryExpModel}
\end{equation}
This allows us to convert the memory term, which was previously an integral, into a dynamic state of the agent. Eq~\ref{eqn:MemoryExpModel} describes the evolution of this memory-state of the agent: $\mathbf{v}_0 - \mathbf{v}_i$ serves as the instantaneous source for the memory while $-\frac{\mathbf{M}_i}{\Tilde{\alpha}}$ is the rate at which memory decays.

\section{Results and Discussion}
\subsection{Dynamics of a single agent}
The equations for the  dynamics of an isolated agent, far away from the boundary, can be written compactly as in Eq~\ref{eqn:singleagent_goveq}. This is after:
\begin{inlineenum}
    \item the governing equations in Eq \ref{eqn:socialforcemodel} are scaled ($t$ with $\tau$, $\|\mathbf{v}\|$ with $\|\mathbf{v}_0\|$, $\|\mathbf{M}\|$ with $\|\mathbf{v}_0\|/\tau$),
    \item the interaction terms are dropped and,
    \item the scaled memory parameters become $\alpha = \Tilde{\alpha}/\tau$ and $\beta = \Tilde{\beta}/m$.
\end{inlineenum}

\begin{equation}
\frac{d}{dt}\begin{pmatrix}
v_x\\ 
v_y\\ 
M_x\\ 
M_y
\end{pmatrix}
=
\begin{pmatrix}
-1 & 0 & \beta & 0\\ 
0 & -1 & 0 & \beta\\ 
-1 & 0 & -\frac{1}{\alpha} & 0\\ 
0 & -1 & 0 & -\frac{1}{\alpha}
\end{pmatrix}
\times
\begin{pmatrix}
v_x\\ 
v_y\\ 
M_x\\ 
M_y
\end{pmatrix}
+
\begin{pmatrix}
1\\ 
0\\ 
1\\ 
0
\end{pmatrix}
\label{eqn:singleagent_goveq}
\end{equation}

\begin{figure*}
    \centering
    \includegraphics[width=0.8\linewidth]{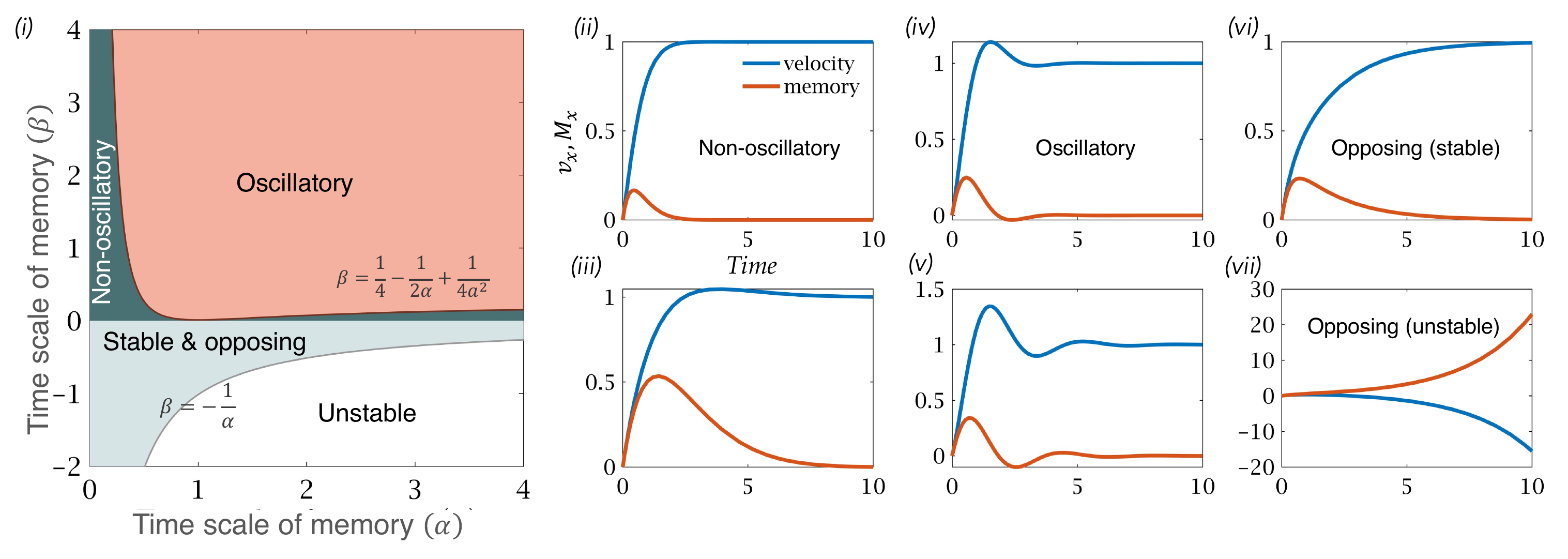}
    \caption{\textit{(i)} Regions with qualitatively different dynamics in the $\beta-\alpha$ parameter space. Coloured (shaded) regions exhibit stable dynamics, which can be oscillatory, non-oscillatory and opposing (negative $\beta$). The unshaded region marks the unstable region, which is found only when $\beta$ takes a sufficiently large negative value. \textit{(ii)-(vii)} Time evolution of the x components of velocity and memory of the isolated agent for different memory parameters. \textit{(ii)} $\alpha=0.3;\beta=2$, \textit{(iii)} $\alpha=3;\beta=0.2$, \textit{(iv)} $\alpha=0.75;\beta=3$, \textit{(v)} $\alpha=3;\beta=3$, \textit{(vi)} $\alpha=0.4;\beta=-1$, \textit{(vii)} $\alpha=3;\beta=-1$.}
    \label{fig:isolated_agent_dynamics}
\end{figure*}

To understand the effect of memory ($\alpha$ and $\beta$) on the movement characteristics of a single agent, it is enough to look at the eigen values of the matrix in Eq \ref{eqn:singleagent_goveq}. The eigen values are $\frac{1}{2\alpha} \times \big[-(\alpha+1) \pm \sqrt{-4\alpha^2\beta + \alpha^2 - 2\alpha + 1}~\big]$, repeated twice.
Analysing the eigen values helps us partition the $\beta-\alpha$ space into four regions (see Fig~\ref{fig:isolated_agent_dynamics}, \textit{i}).

\textbf{Stability:} When $\beta>0$, the memory term aids in restoring the agent to its desired state at steady state (see Fig~\ref{fig:isolated_agent_dynamics}, \textit{ii}). And when $\beta<0$, memory opposes the agent's attempt to reach its desired state. However, since the restitution force $(\mathbf{v}_0 - \mathbf{v}_i)$ always acts to restore the agent's dynamics, a small negative $\beta$ memory does not immediately destabilise the dynamics and only increases the time taken to reach steady state (see Fig~\ref{fig:isolated_agent_dynamics}, \textit{vi}). However, when $\beta<-\frac{1}{\alpha}$, the eigen values begin to have a positive real part and the memory term overpowers the restitution force leading to unstable dynamics (see Fig~\ref{fig:isolated_agent_dynamics}, \textit{vii}).

\textbf{Overshoots \& oscillations:} When $\beta > \frac{1}{4} - \frac{1}{2\alpha} + \frac{1}{4\alpha^2}$, the eigen values become complex conjugates and the dynamics become under-damped which results in oscillations in both the memory and the velocity of the agent (Fig~\ref{fig:isolated_agent_dynamics}, \textit{iv} and \textit{v}).
Also, when $\alpha = 1$, any $\beta$ greater than $0$ results in oscillations. This divides the region exhibiting non-oscillatory dynamics into two. 
The high-$\alpha$ part of this region exhibits dynamics where we observe overshoots; \textit{i.e.} the velocity takes a larger value before asymptotically tapering to the desired, steady state value (see Fig~\ref{fig:isolated_agent_dynamics}, \textit{ii} and \textit{iii} to compare the dynamics in the low-$\alpha$ and high-$\alpha$ regions).

\subsection{Collective escape through a narrow exit}
\begin{figure*}
\centering
    \includegraphics[width=0.8\linewidth]{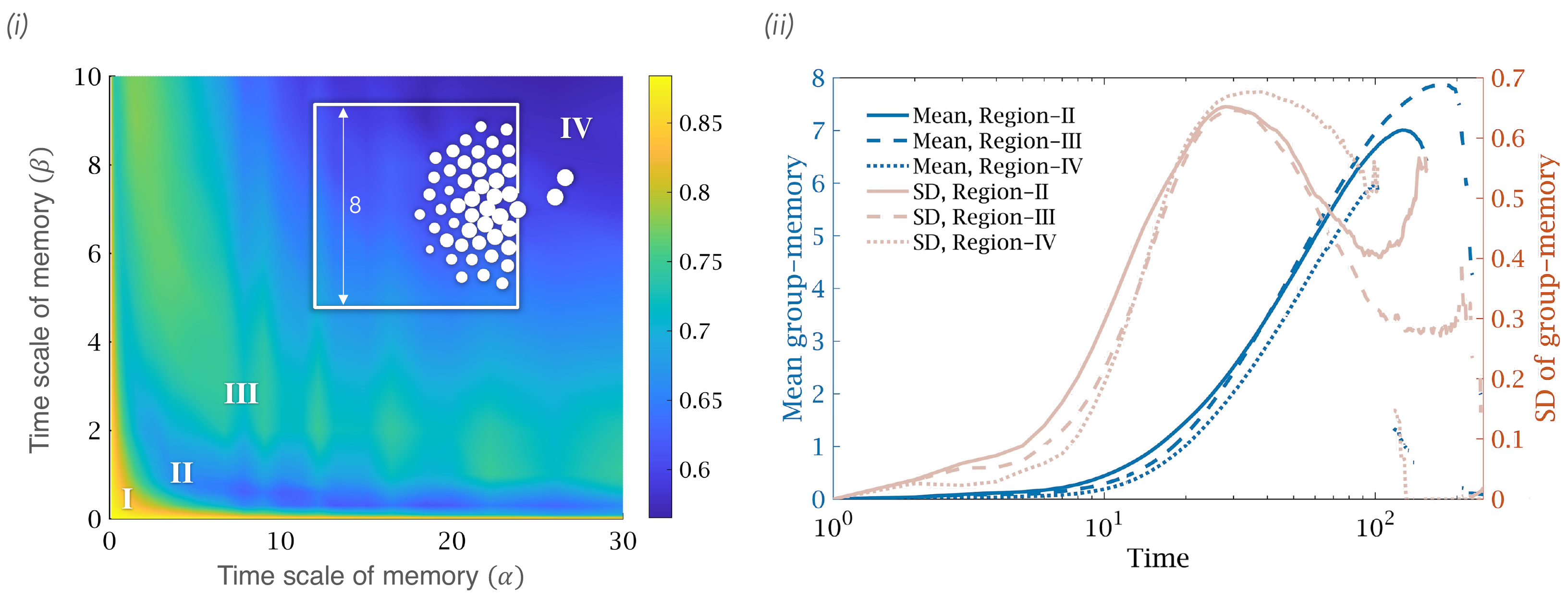}
    \caption{\textit{(i)} - Heat map of the clogging order parameter $\mathcal{C}$, averaged over $75$ independent realisations, plotted for the memory parameters $\beta\in[0, 10]$ and $\alpha\in[0.1, 30]$. INSET: a snapshot of agents exiting a room through a narrow exit. \textit{(ii)} - Time evolution of the mean and the standard deviation of the memory of agents within the room, averaged over $75$ independent realisations. This is plotted for three different memory configurations one from each of the regions II, III and IV. Parameters considered are $\alpha=22.2 \text{ and } \beta=0.3$ for region II, $\alpha=22.2 \text{ and } \beta=2$ for region III and $\alpha=22.2 \text{ and } \beta=10$ for region IV.}
    \label{fig:collectiveescape}
\end{figure*}

To understand the effect of memory on the dynamics of the collective, we consider a well known system: agents exiting a room via a narrow door (See inset of figure~\ref{fig:collectiveescape} \textit{i}). We use the conditions similar to that in ref~\cite{Helbing2000}.
The governing equations in Eq~\ref{eqn:socialforcemodel} and \ref{eqn:MemoryExpModel} are simulated where every agent has a $\mathbf{v}_0$, directed along the line joining the agent centre and the mid-point of the exit of the door. The inter-agent interactions lead to collisions between agents as they crowd near the exit, slowing them down and giving rise to temporary clogging. Now as agents slow down, memory force increases since the source for the memory is $\mathbf{v}_0 - \mathbf{v}_i$. This makes the agents push harder as they attempt to exit the room. To understand how these forces experienced by the agents lead to the collective escape of these agents, we introduce an order parameter $\mathcal{C}$ that characterises the clogging propensity. 
Here, $\mathcal{C}$ is simply the fraction of the total time when the agent-number in the room remains a constant.
With this definition, a high value of $\mathcal{C}$ would correspond to the prevalence of a large number of clogging events during which agent number in the room remains unchanged.

Figure~\ref{fig:collectiveescape} \textit{i}, shows the heat map of $\mathcal{C}$ in the $\beta-\alpha$ space. 
We find the landscape of $\mathcal{C}$ to be non-monotonic with respect to both the principle axes: \textit{i.e.}, an increase in the memory does not necessarily reduce the propensity to form clogs.
This observation is similar to the so-called faster-is-slower effect (FIS), reported by Helbing et al~\cite{Helbing2000,Garcimartin2014, Pastor2015}, where agents trying harder do not necessarily lead to more efficient escape. 
In addition, we find that the observed non-monotonic effect is more pronounced in certain regions of the $\alpha-\beta$ space. For instance, when either $\alpha$ or $\beta$ are high (see figure~\ref{fig:collectiveescape} \textit{i}), the non-monotonicity with respect to the other parameter is well pronounced in comparison to when they are low.
We believe this feature arises due to the intrinsic dependence of the memory force on the parameters $\alpha$ and $\beta$ as seen in figure~\ref{fig:isolated_agent_dynamics}. When an agent remembers only its recent past, \textit{i.e.} when $\alpha$ is small, a high value of $\beta$ is required to produce a force of a similar magnitude to transition from non-oscillatory to oscillatory behaviour and vice versa.

We partition the $\alpha-\beta$ space in figure~\ref{fig:collectiveescape} \textit{i} into four regions while preserving the $\alpha-\beta$ relationship discussed previously. Region I is monotonic with respect to $\mathcal{C}$ and completely covers and extends over the region in the $\alpha-\beta$ space corresponding to non-oscillatory behaviour at the agent-level (compare with figure~\ref{fig:isolated_agent_dynamics}).
Region II corresponds to the ideal amount of memory that agents can possess to efficiently escape through the exit.
Region III corresponds to the FIS phenomenon and it is sandwiched between regions II and IV which exhibit efficient collective escape.

To understand how memory affects the escape dynamics and the formation of clogs, we look at the averaged dynamics of the memory term for the agents within the room. 
Figure~\ref{fig:collectiveescape} \textit{ii} shows the time evolution of the mean and standard deviation of the memory $\|M_i(t)\|$ in system, averaged over all the independent realisations of the escape dynamics. 
The initial transient dynamics show that the effect of memory is more pronounced for the agents with lesser strength of memory (\textit{i.e.}, regions $II>III>IV$). This is because memory increases when the agents begin to slow down and agents with a higher strength of memory reach their desired state faster.
This trend disappears as the agents clog near the entrance: both the mean and standard deviation of memory of the agents within the room increases rapidly in time. The dynamics of agents with memory in region III begin to have a higher overall memory with smaller variation between the agents within the room in comparison to II and IV. In other words, the strength and time scale of memory is such that all the agents try hard to exit by a similar amount which favours the formation of clogs and gives rise to a high value of $\mathcal{C}$. However, the higher variation and the lower mean memory of agents with memory in region IV implies that the some agents have more memory than the rest, which favours a more efficient exit strategy giving rise to lower values for $\mathcal{C}$.

\section{Conclusion}
A key feature of the traffic characteristics of living collectives is the ability of individual agents to assimilate dynamic information and alter their movement appropriately. In this article, we introduce \textit{memory}, a facet of intelligence where an agent remembers its trajectory from its recent past and quantifies how well it was able to achieve its desired velocity. The agent makes up for any non-optimal movement in its recent past with a social force proportional to the memory.
We show that an agent's memory has an effect akin to a Proportional--Integral controller.
% We analyse the effect of the strength of the social force and the time scale of the memory term on the dynamics of an isolated agent detailing the regions in the $\beta-\alpha$ space that give rise to qualitatively different behaviour: presence and absence of oscillations, stable and unstable.
Depending on the memory parameters $\alpha$ and $\beta$, the agent dynamics can be stable or unstable, and give rise to overshoots and oscillations.
The eigen values of the system shed light on the boundaries that partition the different regions in the $\beta-\alpha$ space that show qualitatively different dynamics. 

While the effect of memory on the dynamics of the agent is monotonic when only a single agent is considered, it is not the case when agents are in a crowd. We study the effect of memory on the dynamics of agents exiting a room through a narrow door. We find that the presence of memory does not always improve the movement of the agents: the clogging order parameter $\mathcal{C}$ has a non-monotonic landscape in the $\beta-\alpha$ space resulting in a behaviour similar to the well known faster-is-slower effect. This allows us to partition the $\beta-\alpha$ space into four regions based on a order parameter that quantifies clogging propensity. 
% Region I is where the dynamic behaviour is monotonically becoming more efficient with increase in memory. $II-III-IV$ correspond to the non-monotonic regions in the $\beta-\alpha$ space. 
We find the eigen values corresponding to the motion of individual agents, in the non-monotonic region of the $\beta-\alpha$ space, to be complex. In other words, the observed FIS effect is a result of the under-damped response of agents arising due to the integral component of the memory force term.
The time dynamics of the mean and standard deviation of the memory of agents within the room, reveal why some regions in the $\beta-\alpha$ space favour the formation of clogs giving rise to non-monotonicity.

\vspace{10pt}
\section*{Acknowledgements}
The authors thank the DST INSPIRE faculty award (grant number: DST/INSPIRE/04 /2017/002985) for the funding. DRM thanks V Kumaran, IISc Bangalore for the discussions during the early stages of the work.

\bibliography{references.bib}
\end{document}